\documentclass[a4paper]{article}
\usepackage{graphicx} 
\usepackage{lineno}
\usepackage{subcaption}
\usepackage{gensymb}
\usepackage[english]{babel}
\usepackage{authblk}
\usepackage{cite}

\usepackage{amsmath,amssymb,amsfonts}
\usepackage{algorithmic}
\usepackage{textcomp}
\usepackage[T1]{fontenc}
\usepackage[backref]{hyperref}
\usepackage{mathptmx} 
\usepackage{geometry}
\usepackage{fancyhdr}
\usepackage[utf8]{inputenc}
\usepackage{caption}
\usepackage{abstract}

\title{Demonstration of Photonics-based D-band Integrated Localization and Communication}

\author[1,\dag,*]{Qigejian Wang}
\author[1,\dag]{Yirui Deng}
\author{Deepak Mishra}
\author{Yixuan Xie}
\author{Elias Aboutanios}
\author[1,*]{Shaghik Atakaramians}

\affil[1]{School of Electrical Engineering and Telecommunications, UNSW Sydney, NSW 2052, Australia.}
\affil[$\dag$]{These authors contributed equally to this work.}
\affil[*]{qigejian.wang@student.unsw.edu.au, s.atakaramians@unsw.edu.au}

\date{}

\begin{document}

\maketitle

\begin{abstract}
The Terahertz spectrum has the ability to provide high-speed communication and millimeter-level resolution. As a result, terahertz-integrated sensing and communication (ISAC) has been identified as a key enabler for 6G wireless networks. This work discusses a photonics-based D-band communication system for integrated high-resolution localization and high-speed wireless communication. Our empirical results show that a communication rate of 5 Gbps over a distance of 1.5 meters and location identification of the target with millimeter-level ($< 3$~mm) range resolution can be conducted simutaneously. We also show that the error due to the thickness of the beam splitter can be eliminated, while the quantization error and the random drift errors are the limiting factors of the resolution achieved. This experimental demonstration using D-band communication indicates that terahertz ISAC can be realized for 6G networks while considering the underlying system restrictions, e.g. bandwidth limit and lens diameter. 
\end{abstract}

\section{Introduction}
Sixth-generation (6G) wireless technology has promised to deliver a peak data rate of 1 terabits/s with less than 100 microseconds latency~\cite{nagatsuma2016advances,liu2023closed}. Terahertz spectrum has been identified as one of the enablers for the realization of 6G broadband connectivity~\cite{Mobiledata,SKTelecom}. Attenuation of terahertz radiation by atmospheric gases such as water vapor and oxygen molecules limits terahertz transmission to specific frequency windows, e.g. D-band ($110-170$ GHz), H-band ($220-325$ GHz), and Y-band ($325-500$ GHz)~\cite{ma2018invited,gil2021estimation,nagatsuma2016advances}, which have relatively low atmospheric absorption, leading to the potential for high-speed communication with hundred GHz of bandwidth and pico-second-level symbol duration~\cite{niu2015survey,nagatsuma2016advances,petrov2018last}.

Recently integrated localization and communication have gained momentum. Localization is the process of estimating the position of the target and if required the orientation, which is essential for location-aware communications~\cite{lemic2022toward,kanhere2021outdoor} and tactile internet~\cite{promwongsa2020comprehensive,hou2021intelligent}. Using the current wireless network, localization and tracking have been demonstrated in outdoor scenarios such as vehicles on the road with meter-level accuracy using a carrier frequency of 28 GHz~\cite{sun2020fingerprinting}, and indoor scenarios using WiFi at 2.4 GHz and 5 GHz band~\cite{guan2023experimental}. In addition, positioning with relative lateral and longitudinal accuracy of $0.1$ m and less than $0.5$ m, respectively, has been demonstrated using the 5G network for self-driving vehicles~\cite{wang2023recent}. However, this is still not accurate enough to support applications such as future assistant robots in smart factories, where high precision in centimetre level (< $2$ cm) is needed for opening doors and picking up items~\cite{tan2021integrated}. Radar function has also been integrated with existing communication systems, IEEE 802.11 (including but not limited to 2.4 GHz, 5 GHz, 6 GHz, and 60 GHz) frequency bands~\cite{kumari2017ieee,daniels2017forward,sit2011doppler,nguyen2017delay} and 3rd Generation Partnership Project (3GPP, also known as 5G New Radio) frequency bands~\cite{barneto2019full,kanhere2021target,petrov2019v2x}, which can only provide centimeter-level resolution due to limitation on bandwidth and data rate.

In future 6G networks, integrated localization and communication is vital to achieve ubiquitous connectivity with high data rates and low latency. Terahertz band can offer improved localization performance due to large bandwidth availability and short wavelengths. Heretofore, there is some research work on the integration of imaging within terahertz communication systems~\cite{li2021integrated,li2021integrated2}, while integrated localization and communication are yet to be widely investigated for deployment in the terahertz band. Moreover, most of limited existing works use conventional frequency modulated signal for target detection, which has challenges including the requirement of an additional waveform apart from the communication signals, relatively low range resolution, and two functions not working simultaneously. For example, a range resolution of $1.58$ cm has been demonstrated using linear frequency modulation (LFM) for radar, and orthogonal frequency-division multiplexing (OFDM) signals in a 340 GHz (Y-band) photonics-based communication system~\cite{wang2022integrated}. In terms of D-band, a joint radar-communication complementary metal–oxide–semiconductor (CMOS) transceiver using an electronics-based method has been demonstrated, achieving a 1.25 cm range resolution with frequency-modulated continuous-wave (FMCW) signal at 150 GHz by switching between communication and radar mode \cite{deng2022d}. To our best knowledge, simultaneous localization and communication system with millimetre-order resolution at D-band has not yet been reported.

Here, we demonstrate integrated high-resolution localization using a D-band photonics-based communication system. Millimetre order accuracy ($< 3$ mm) can be achieved using the same system simultaneously. The content is organized as follows: first, we describe the D-band communication system with 5 Gbps data rate and 1.5 m link distance, and investigate the system's capabilities and the hardware limitation for the communication link. We then utilize a squared-wave signal for measuring the distance of the target, followed by the discussion of detection accuracy and resolution.

\section{Terahertz communication system}
Currently two approaches are utilized for realizing terahertz communication systems: photonics-based and electronics-based~\cite{petrov2018last}. Electronics-based systems use frequency multiplying chains to up-convert a lower-frequency signal from the microwave and millimeter-wave bands~\cite{song201450,rodriguez2020qpsk,gil2021estimation}. These systems have a simple setup and high terahertz power (order of milliwatt~\cite{kasagi2019large,biswas20181} and watt~\cite{baig2017performance,hu2019demonstration}). However, the electronics-based systems suffer from waveform distortion and high phase noise due to non-linear effects of frequency multipliers and mixers. The high gain usually leads to reduction of the bandwidth~\cite{akyildiz2022terahertz}.

The photonics-based systems use photomixing to achieve optical-to-terahertz down-conversion. Two laser beams are used to generate the terahertz beat note~\cite{nallappan2018live,stohr2017coherent,kato2021photonics}. Photonics-based systems come with much lower terahertz power than the electronics-based systems ($<10\%$). Nevertheless, terahertz signals generated from photonics-based systems have higher spectral purity and lower amplitude and phase noise compared to electronics-based systems. The existing optical modulators can provide high bandwidth and modulation index. Furthermore, the photonics-based systems have the potential to be integrated in existing optical-fiber networks and harness the photonics advances for miniaturization of hardware~\cite{richardson2013space,akyildiz2022terahertz}.

Terahertz communication links using electronics-based~\cite{hirata20105,wang201310,abu2021end,schultze2022observations,vassilev2018spectrum,ando2016wireless} and photonics-based~\cite{rodrigues2020hybrid,nallappan2018live,wang2022complex}, approaches have been demonstrated for D-band. This while mainly photonics-based communication systems has been demonstrated for H-band~\cite{castro201932,maekawa2023300,su2021demonstration,kato2021photonics,song201450,antes2012220,rodriguez2020qpsk} and Y-band~\cite{wang202026,ding2022104,jia2022integrated,wang2023demonstration,ma2018invited}.

\subsection{Photonics-based terahertz communication system}
\begin{figure}[htbp]
\centering\includegraphics[width=10cm]{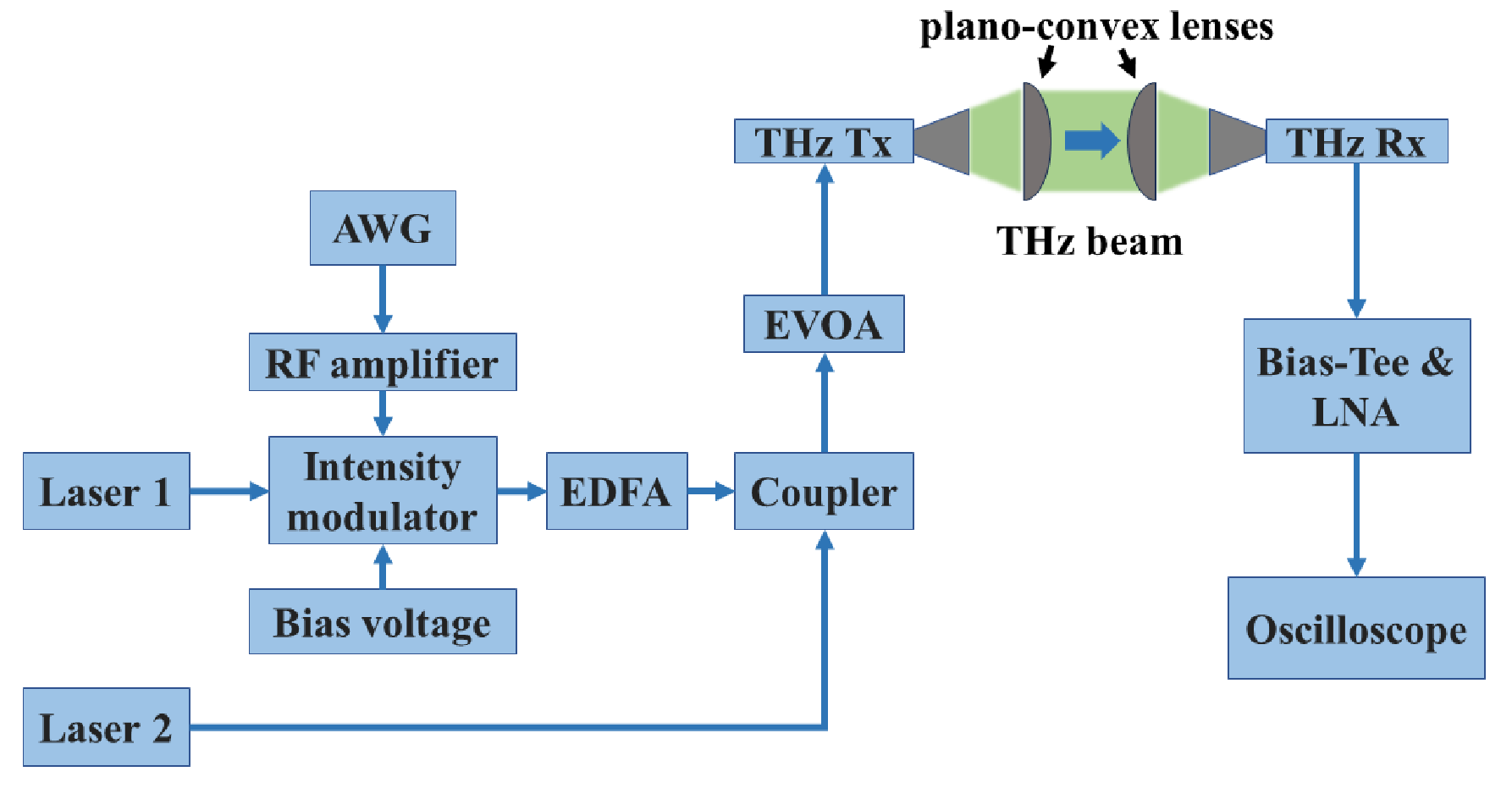}
\caption{Schematic of the photonics-based terahertz wireless communication system with modulation on one laser. (AWG: Arbitrary waveform generator, EDFA: Erbium doped fiber amplifier, EVOA: Electronic variable optical attenuator, LNA: Low noise amplifier).} \label{setup}
\end{figure}

Figure~\ref{setup} shows the schematic of the photonics-based terahertz system. The system consists of two tunable 1550 nm lasers (Toptica DFB with 1 MHz resolution). The lasers operate at the terahertz difference frequency, in this case D-band. The baseband signal is modulated on one laser before coupling with the second laser (Fig.~\ref{setup}), providing a single sub-carrier modulation. The modulation can also happen after the coupling of two lasers, providing a double sub-carrier modulation, which is claimed to have higher terahertz signal-to-noise ratio (SNR)~\cite{nagatsuma2013terahertz}.

\begin{figure}[htbp]
\centering\includegraphics[width=10cm]{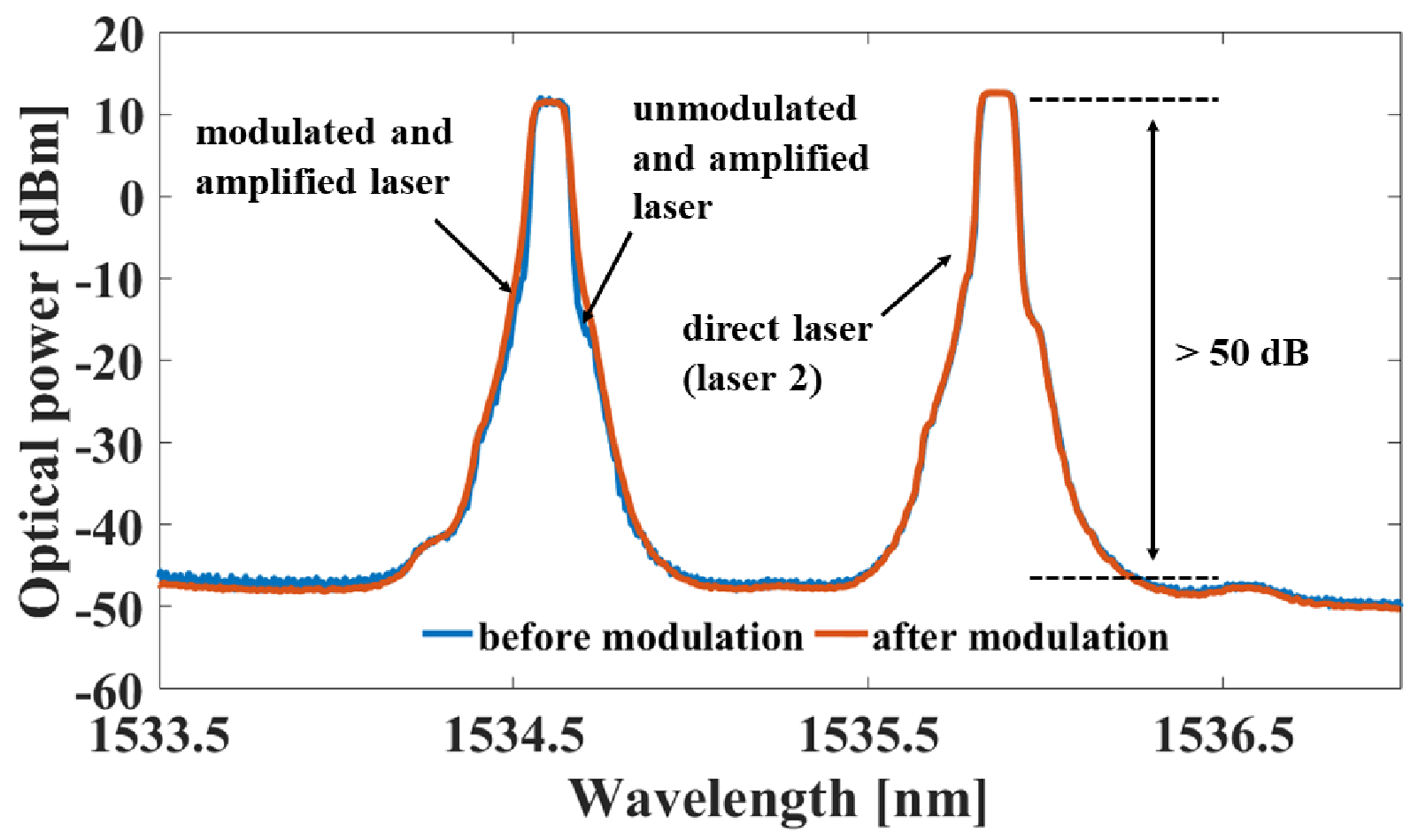}
\caption{Optical spectrum of the optical signals with frequency separation of 160 GHz for terahertz wave generation, where the blue curve is before modulation and the red curve is after modulation.} \label{osa}
\end{figure}

An arbitrary waveform generator (AWG, Keysight M8190A) is used to generate the baseband signal with 400 m$V_\mathrm{pp}$ amplitude. The generated RF signal is first amplified to 5.5 V by a modulator driver (Thorlabs MX40A) and then modulated (amplitude shift keying modulation, ASK) to one laser using an electro-optic Mach-Zehnder modulator (Thorlabs LN05S-FC). The modulator driver has a lower cutoff frequency of 100 KHz, which as demonstrated later leads to error in modulation of signals with low frequencies, e.g. acoustic signals. The bias voltage of the modulator is controlled by the modulator driver, which continuously monitors quadrature bias point with positive slope using a dither tone (1 KHz, 600 m$V_\mathrm{pp}$) to correct drift in the bias voltage. Since the output optical power of the modulator is low (< 2 mW), an erbium doped fiber amplifier (EDFA, Thorlabs EDFA100P) is used after the modulator to amplify the laser power to $\sim 35$ mW, similar power level to the unmodulated laser 2. A 3 dB coupler is used to combine the lasers. An electronic variable optical attenuator (EVOA, Thorlabs EVOA1550A) is connected before the terahertz transmitter to limit the optical power to avoid overloading the photomixer. Figure~\ref{osa} shows the optical spectrum of the two lasers with a frequency separation of 160 GHz measured using an optical spectrum analyzer, where the baseband signal is modulated on laser 1. Slight increase in the bandwidth is observed after modulation (red curve), comparing to the optical signal without modulation (blue curve). The signal-to-noise ratio (SNR) is higher than 50 dB for both modulated and unmodulated optical signals.

The terahertz beam is generated by feeding the combined laser into either a D-band uni-traveling-carrier-photodiode (UTC-PD) photomixer (NTT D-band photomixer module) or a broadband antenna-integrated InGaAs photodiode (Toptica TeraScan 1550). The UTC-PD can provide $\sim 0.25$ mW of terahertz power through D-band, while the InGaAs photodiode can provide $\sim 0.1$ mW in the same band. A horn antenna with 25 dBi gain is connected to the UTC-PD to couple the generated beam into free space, this is while a silicon lens is used for InGaAs photodiode.

For free-space terahertz transmission, two plano-convex lenses with focal length of 50 mm and diameter of 35 mm are used to collimate the terahertz beam to free space and focus it into the detector (Fig.~\ref{setup}). On the receiver side, another horn antenna is used to couple the signal to a D-band zero-bias Schottky diode (ZBD, Virginia Diodes WR6.5ZBD-F20), which also demodulate the terahertz signal. The DC component in the demodulated baseband signal is then filtered using a Bias-Tee and amplified using a low noise amplifier (LNA) with $\sim 14$ dB gain. The amplified signal is finally measured by a real-time oscilloscope (Keysight Infiniium-MXR-Series, with 6 GHz bandwidth).

\subsection{Characterization of the terahertz communication system}
Here, first we investigate the effect of atmospheric humidity on the D-band communication link. Then we demonstrate a 5 Gbps communication link for up to 1.5 m distance. We discuss the current limitation of the system in terms of
maximum link distance and the base-band signal and discuss hardware changes for improvements. 

\subsubsection{Effect of humidity on communication band}

\begin{figure}[htbp]
\centering\includegraphics[width=10cm]{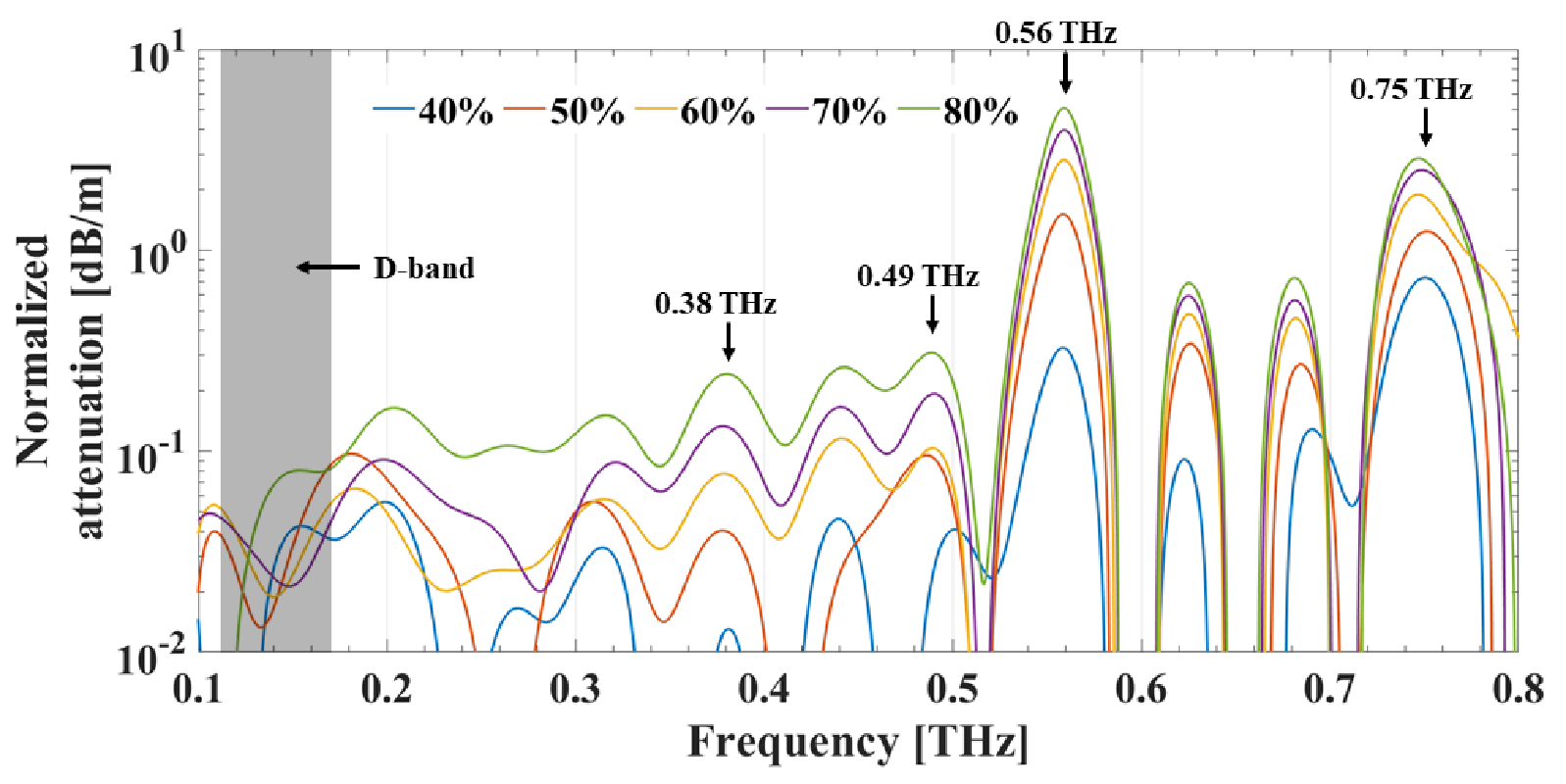}
\caption{Measured normalized attenuation through a cubic acrylic box filled with air with humidity changing from $40\%$ to $80\%$.} \label{humidity}
\end{figure}

Atmospheric humidity absorption is one of the key factors which can affect the communication quality at higher frequencies. To study the terahertz absorption from water molecules in the air for short link distance, we conducted an experiment measuring the terahertz transmittance of an enclosed cubic acrylic box filled with air, where the humidity was adjusted from $40\%$ to $80\%$ using a humidifier. The terahertz transmitter and receiver were set close to two opposite sides of the box with a distance of 55 cm. Figure~\ref{humidity} shows the measured attenuation normalized to the original room humidity ($37\%$) at 0.1 - 0.8 THz. As expected from existing literature~\cite{ma2018invited, slocum2013atmospheric}, there is minimal effect on transmission due to the humidity increases at D-band. Significant absorption appears at 0.56 and 0.75 THz, which are the dominated humidity absorption points also reported in literature~\cite{ma2018invited, slocum2013atmospheric}. Moreover, absorption peaks appear at other frequencies $\sim$0.38 THz and $\sim$0.49 THz are also reported in literature, which are due to other gas composition of the atmosphere such as oxygen~\cite{slocum2013atmospheric, series2019attenuation}.

\subsubsection{Bit error rate measurements}

\begin{figure}[htbp]
\centering\includegraphics[width=10cm]{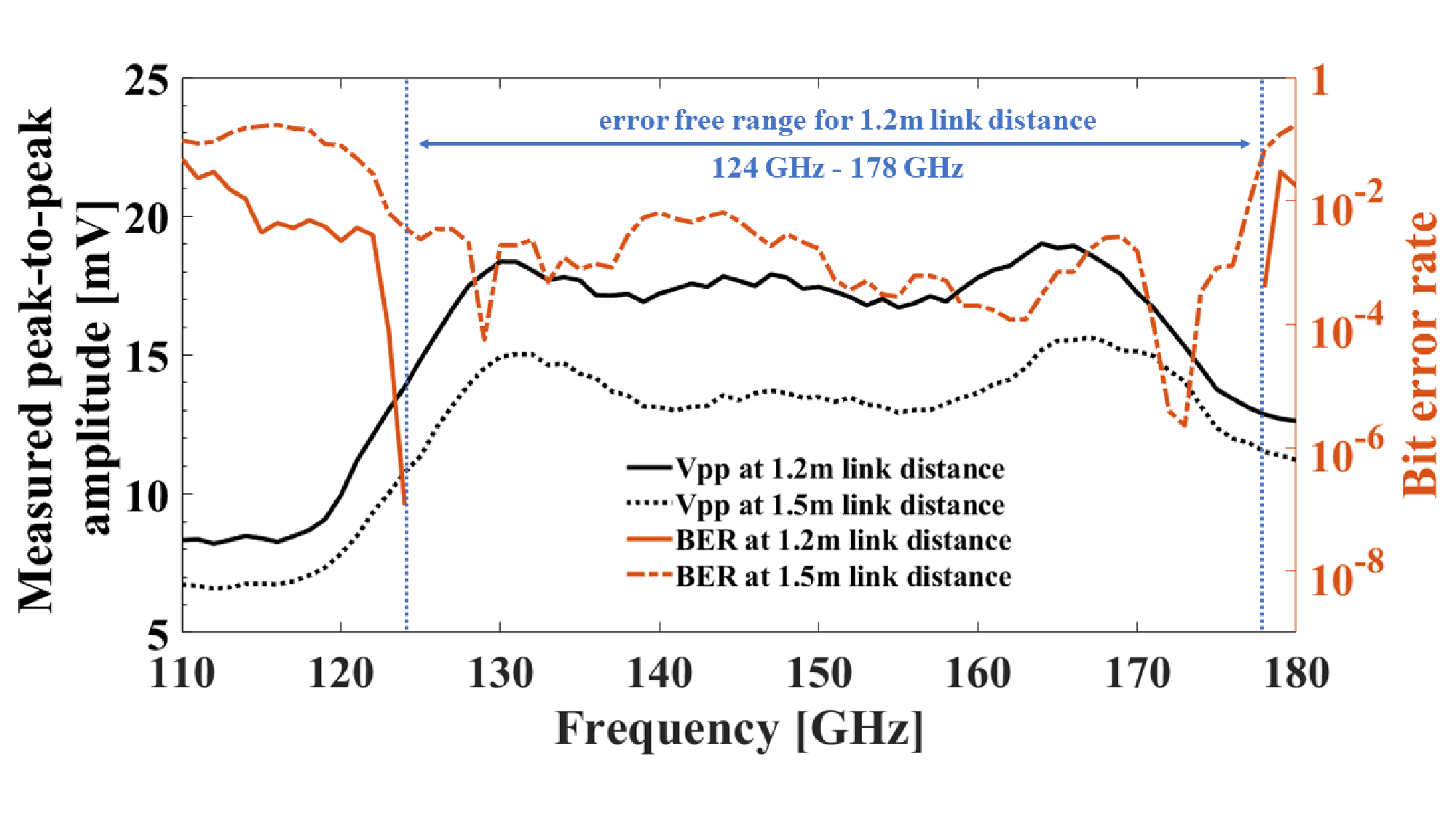}
\caption{Measured $V_\mathrm{pp}$ and BER of a 5 Gbps PRBS baseband signal with carrier frequency of $110-180$ GHz for 1.2 m (solid lines) and 1.5 m (dashed lines) link distances.} \label{ber}
\end{figure}

Received peak-to-peak amplitude ($V_\mathrm{pp}$) and bit error rate (BER) are two key parameters for communication quality evaluation, which can be measured by the oscilloscope. To characterize the system performance, here, we used a Non-Return to Zero (NRZ) pseudo random binary sequence (PRBS) data with 5 Gbps bit rate and 400 m$V_\mathrm{pp}$ amplitude as the baseband signal (amplitude modulation). First we measured the $V_\mathrm{pp}$ and BER for carrier frequency from 110 GHz to 180 GHz at link distance of 1.2 m and 1.5 m, shown in Fig.~\ref{ber}. The bias voltage and photocurrent on the terahertz transmitter are fixed at -2 V and 7 mA, respectively, which are recommended values by manufacturer. It can be observed that 130 - 165 GHz frequency range has high measured $V_\mathrm{pp}$ and low BER, which is consistent with the output and responsivity characterization of the UTC-PD and ZBD, respectively. There is an error-free transmission window, 124 - 178 GHz, for 1.2 m link distance, while for 1.5 m link distance there is no error-free range and the overall received $V_\mathrm{pp}$ drops more than 5~mV. We attribute the increase in BER for longer link distance to the divergence of terahertz beam, which is partially collected using 35~mm lenses. The diameter of the terahertz beam, $D_\mathrm{T}$, is a function of the link distance, which can be approximated as $D_\mathrm{T} \sim L \cdot \lambda /D_\mathrm{L}$, where $L$ is the link distance, $\lambda$ is the wavelength of the carrier wave, and $D_\mathrm{L}$ is the diameter of the collimating lenses~\cite{nallappan2018live}. Therefore, in our case, the terahertz beam diameter at the receiver is 93.6 - 60.3 mm for 1.2 m and 117 - 75.4 mm for 1.5 m through D-band, which are more than two times larger than the diameter of the lens (35 mm), leading to only partial signals being received by the terahertz detector. In terms of received terahertz power, if using two identical lenses to collimate the beam, the received power is $P=P_\mathrm{0} \cdot (D_\mathrm{L}/D_\mathrm{T})^2 $~\cite{nallappan2018live}, where $P_\mathrm{0}$ is the emitted power from the terahertz transmitter. Therefore only $13\% - 34\%$ and $9\% - 22\%$ of the emitted power is received at 1.2~m and 1.5~m link distance, respectively. It can be seen that in order to increase the ratio of received power, using lenses with larger diameter is a solution. For example, if lenses in the system are replaced by 100~mm diameter lenses, the system is expected to receive all of the emitted terahertz signal for < 3.6~m link distances without other modification of the system. In addition, link distance can be extended by applying terahertz LNAs. If using a terahertz LNA with 30~dB gain (commercially available), the link distance could be extended to 64.8~m with 42~cm diameter lenses.

\subsubsection{Transmission of low-frequency signal} 

\begin{figure}[htbp]
\centering\includegraphics[width=10cm]{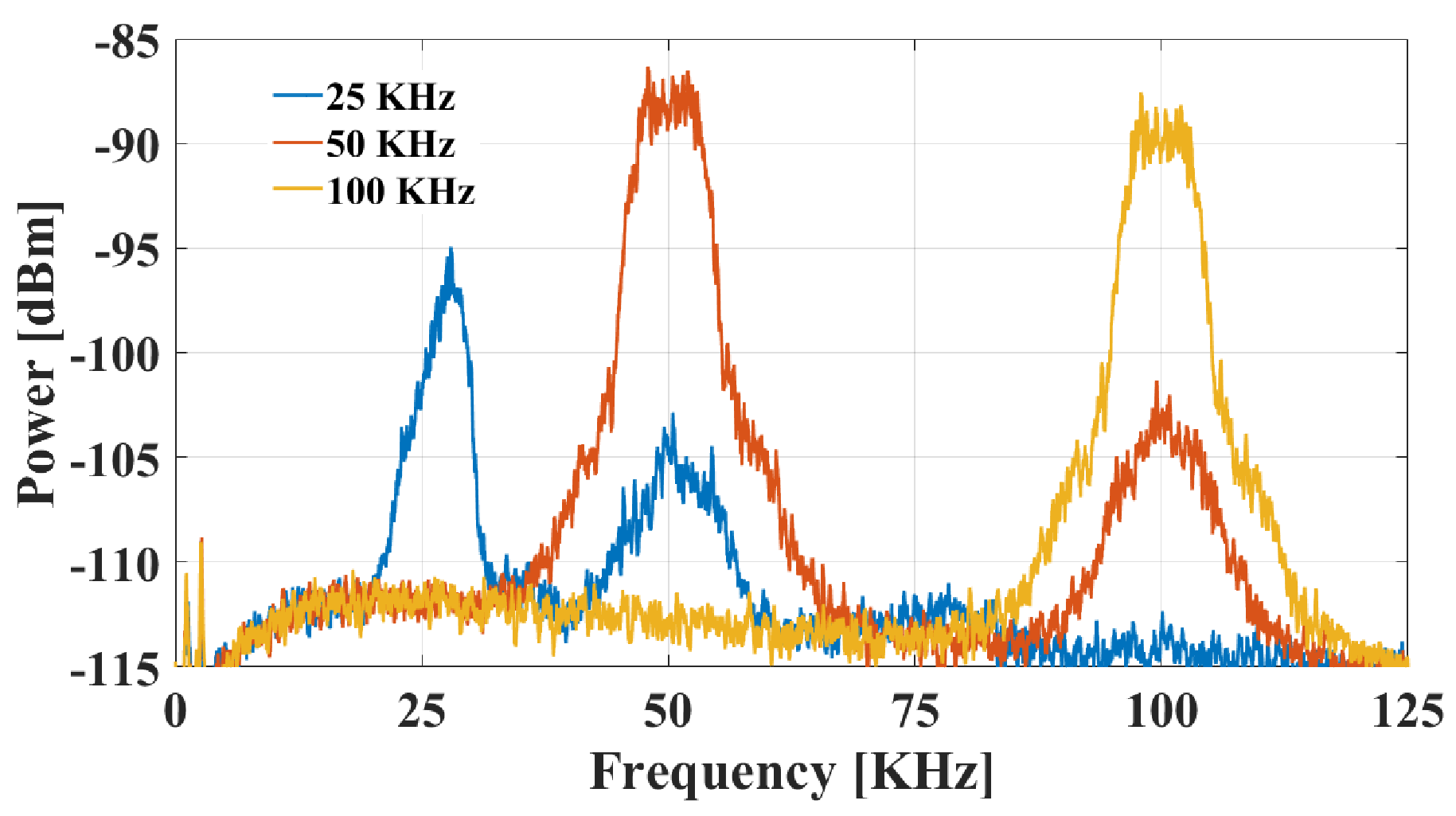}
\caption{Measured spectrum of transmitted power of baseband signals with centre frequency of 25 KHz, 50KHz, and 100KHz with 8 KHz half power bandwidth.} \label{acoustic}
\end{figure}

To understand the limitation imposed by the modulator driver on the baseband signal, we have explored modulating wavefomrs with 8 KHz half power bandwidth with center frequency of 100 KHz, 50 KHz and 25 KHz. The measured waveforms at oscilloscope, Fig.~\ref{acoustic}, confirms undistorted transmission of 100 KHz and 50 KHz signals, while the the 25 KHz signal is not transmitted in full. In theory, we expect to modulate any waveforms including acoustic, but in principal we are limited with the cut-off frequency of the modulator driver.

\section{Target localization using the communication system}
\label{sec_distance}
Here we investigate the accuracy we can achieve to detect the location of a target within the communication channel. For this purpose we send a simple waveform and determine the distance of the target from the reflected time of flight (ToF), a technique widely used for distance measurement~\cite{lee2010time, kolb2010time}.

\subsection{Experimental setup for localization of target}
\label{subsec_dist_setup}

\begin{figure}[t!]
\centering\includegraphics[width=0.7\linewidth]{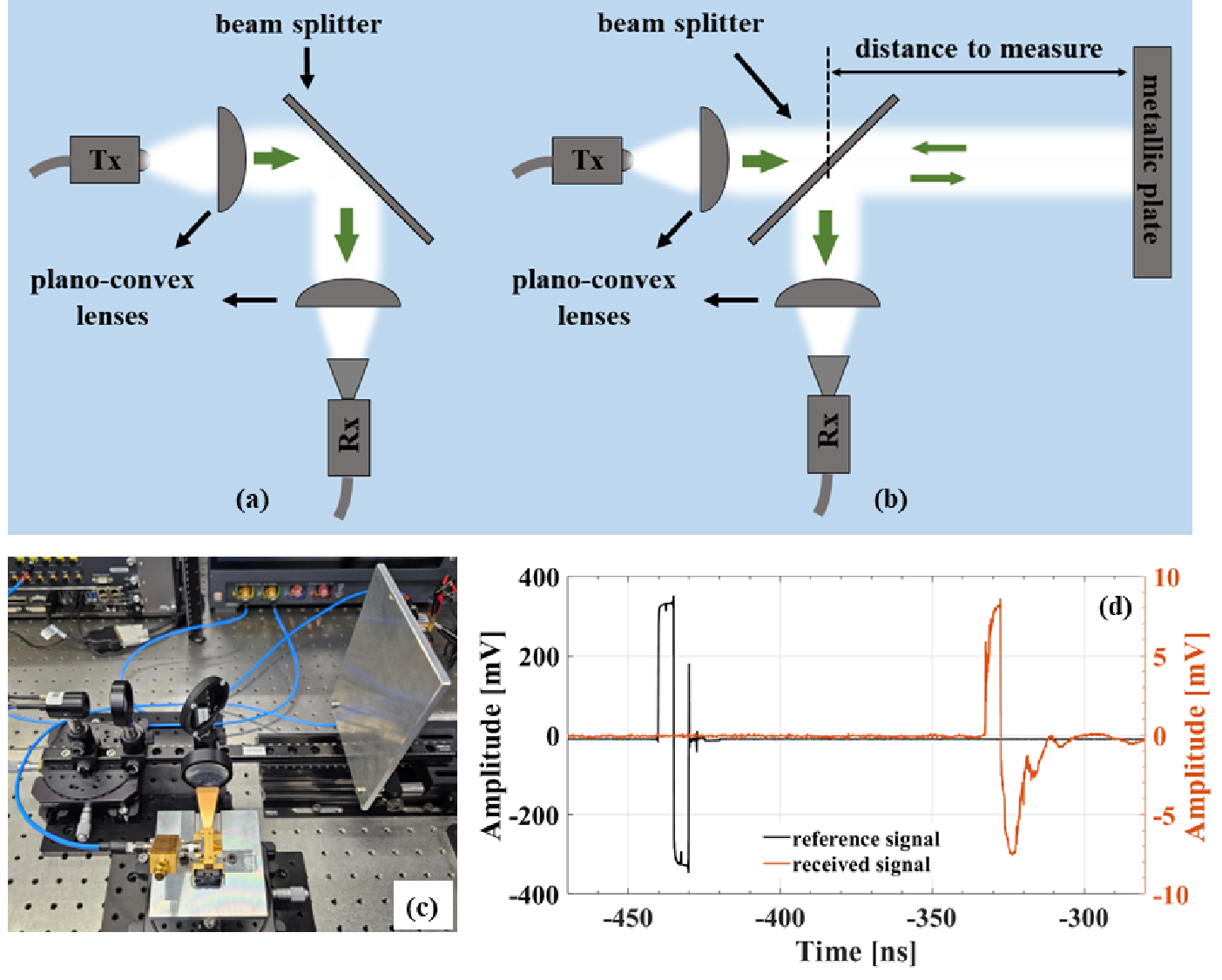}
\caption{Schematic of the experimental setup for reflection measurements, where Tx is the transmitter and Rx is the receiver. (a) Direct reflection for measuring system delay. (b) Normal incident reflection for measuring the distance between the system and a metallic plate. (c) A photograph of the setup of normal incident reflection for distance measurement. (d) Waveforms used for measurements, where the reference signal is the direct output from AWG and the received signal is from the terahertz receiver.} \label{setup_refl}
\end{figure}

The schematic of the experimental setup is shown in Fig.~\ref{setup_refl} (a)-(b), where the arrangement in Fig.~\ref{setup_refl} (b) is commonly used for terahertz reflection spectroscopy~\cite{headland2015terahertz,wang202320,thigale2023terahertz}. A silicon beam splitter is used to split and reflect the terahertz beam, which can be rotated along the vertical axis without changing the position. A metallic plate is set as the target, which is perpendicular to the output terahertz beam from the beam splitter. Figure~\ref{setup_refl} (c) shows the photograph of the experimental setup, where the target distance is the distance between the beam splitter and the metallic plate. The transmitted signal is a single-cycle squared waveform with a cycle length of 10 ns, which is generated by the AWG every 400 ns. Figure~\ref{setup_refl} (d) shows the reference and received waveforms when the system is arranged as shown in Fig.~\ref{setup_refl} (b), where the reference signal is the direct output from AWG and the received signal is from the terahertz receiver. With extracted measurement data, the time delay between the two waveforms is calculated using the cross-correlation method in a back-end processor (Matlab). Meanwhile, the time delay introduced by the terahertz system is measured using the setup shown in Fig.~\ref{setup_refl} (a), where the beam splitter is rotated for $90\degree$, directly reflecting the signal from the terahertz transmitter to the receiver. Intuitively, the total measured delay ($\tau_{\mathrm{m}}$) is the summation of the system delay ($\tau_{\mathrm{sys}}$) and the ToF for the round trip of the target distance. Thus, the estimated target distance $\hat{d}$ can then be calculated with the following equation:
\begin{equation}
\label{eqn_distance}
    \hat{d} = \frac{1}{2}(\tau_{\mathrm{m}}-\tau_{\mathrm{sys}})*c,
\end{equation}
where $c$ is the speed of light, $\tau_{\mathrm{m}}$ denotes the measured total delay of the received signal against the transmitted signal, and $\tau_{\mathrm{sys}}$ denotes the system delay acquired as shown in Fig.~\ref{setup_refl} (a).

Note that since the calculations for both $\tau_{\mathrm{m}}$ and $\tau_{\mathrm{sys}}$ are performed with the discrete data samples extracted from the oscilloscope, the step size of the cross-correlation is the sampling period of the data. Thus, the distance detection resolution of this system, in terms of the minimum recognizable distance change, is constrained by the sampling resolution. For data with a sampling rate $f_\mathrm{q}$, the expected sampling resolution of the system for distance detection is $\Delta d_\mathrm{Q} = \frac{c}{f_\mathrm{q}}$. Particularly in this system, the oscilloscope provides a sampling frequency of $16$~GHz with an $8\times$ interpolation, resulting in an effective sampling resolution $\Delta d_\mathrm{Q} \approx 1.17$mm. Since the output of the system can only be discrete values with $1/f_\mathrm{q}$ steps, any distance $d$ has a quantization error $e_\mathrm{q}$, which is the deviation to its nearest quantization level that,
\begin{equation}
\label{eqn_qerr}
    e_\mathrm{q} = 
    \begin{cases}
    d \;\text{mod}\; \Delta d_\mathrm{Q} & (k\Delta d_\mathrm{Q}\leq d \leq (k+0.5) \Delta d_\mathrm{Q},\; k \in \mathbb{Z})\\
    \Delta d_\mathrm{Q} - d \;\text{mod}\; \Delta d_\mathrm{Q} & ((k+0.5)\Delta d_\mathrm{Q}\leq d \leq (k+1) \Delta d_\mathrm{Q},\; k \in \mathbb{Z}) \\
    \end{cases}.
\end{equation}

We took measurements with terahertz transmitter and receiver, lenses and beam splitter remaining at their respective fixed locations, while the metallic plate (target) moved to various locations (closer or further) along a fixed guide rail. The orientation of the metallic plate remained unchanged. All measurements were taken when the metallic plate was steady for zero-doppler distance. The ruler on the guide rail provided the ground-truth distances for each measurement, while the distance between the beam splitter and the zero point of the ruler was pre-measured as a reference. We carried out multiple rounds of trials, where the metallic plate was randomly moved to $20$ to $25$ different locations on the guide rail with ground-truth distances from $45$~mm to $120$~mm.

\subsection{Distance Detection Outcomes}
\label{subsec_dist_result}

\begin{figure}[!ht]
\centering
    \begin{subfigure}{0.48\textwidth}
    \centering\includegraphics[width=6.2cm]{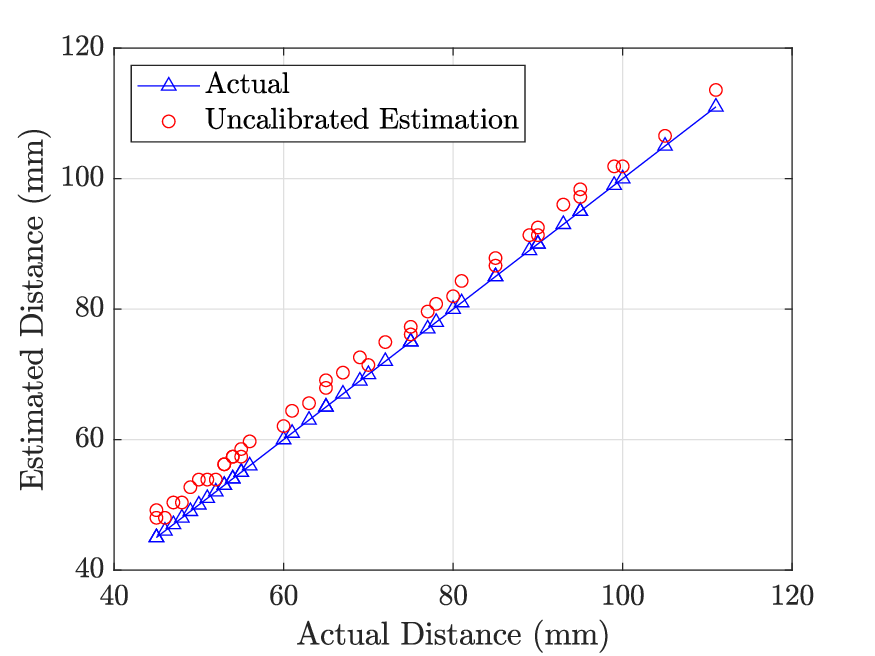}
    \caption{ } \label{result_dist}
    \end{subfigure}
    \begin{subfigure}{0.48\textwidth}
    \centering\includegraphics[width=6.2cm]{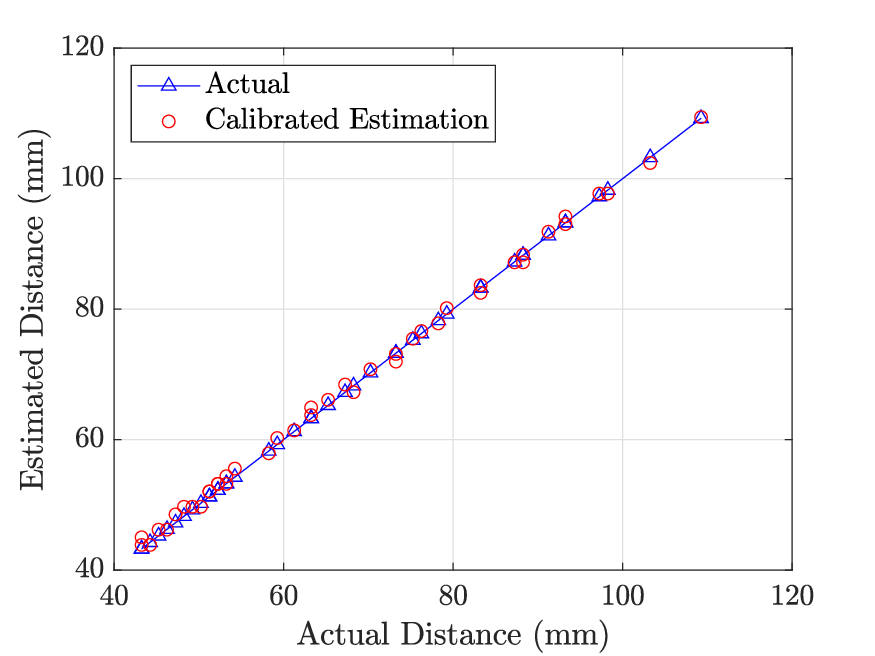}
    \caption{ } \label{result_cal}
    \end{subfigure}
    \caption{Distance estimation outcomes (a) before calibration and (b) after calibration.}
    \label{result_total}
\end{figure}

With all measurements from multiple experimental trials, the outcome of the ToF distance estimation is presented in Fig.~\ref{result_total}(a). The horizontal axis shows the ground truth distances of the target locations, while the vertical axis indicates the estimated distances from the terahertz communication system using Eq.~(\ref{eqn_distance}). The red circles represent the estimated distance value for each measurement, while the blue triangles mark the corresponding ground-truth values. It is observable that the estimated distances are close to the actual distance, and show the same linear increasing trend validating the feasibility of such a distance detection using the D-band communication system. However, a $3$~mm to $6$~mm error exists between the estimated distances and the respective ground truth. We next discuss the potential sources of these errors and system calibration. 

\subsubsection{Distance Detection Accuracy and Calibration}
\label{subsec_calibration}
As discussed earlier, we expect the system to deliver distance estimations with a resolution close to the hardware boundary $\Delta d_{\mathrm{Q}} \approx 1.17$mm. A drifting error of $3$ mm to $6$ mm observed is severe in terms of accuracy performance to this scale of sensitivity. Therefore, we took a deep dive into the estimation errors presented by the system.

\begin{figure}[htbp]
\centering\includegraphics[width=9cm]{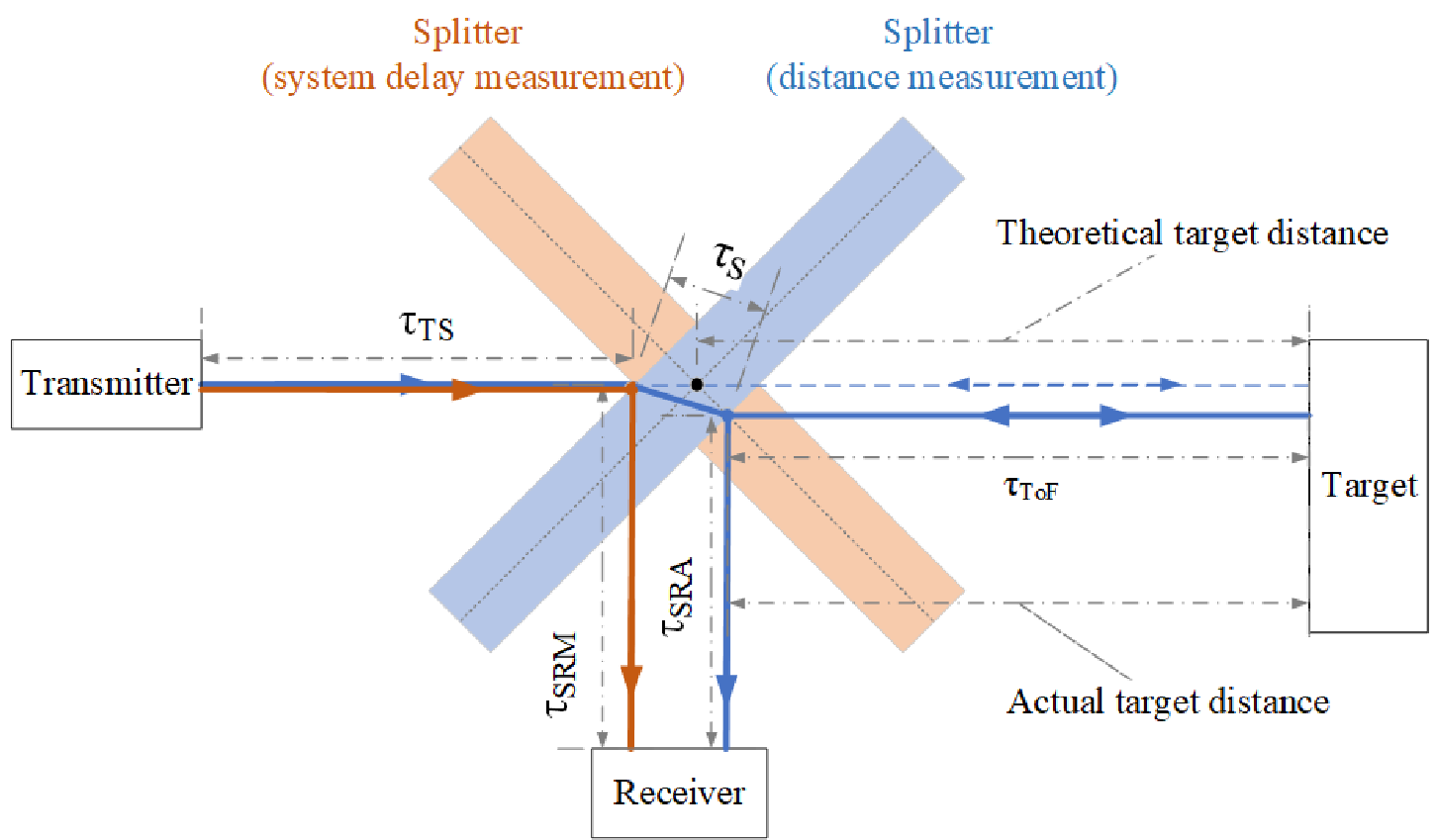}
\caption{Impact of non-ideal beam splitter on system delay measurement and ToF calculation.} \label{result_split}
\end{figure}

Firstly, we noticed that there was a relatively constant offset due to the imperfect ground truth reference or system delay computation. One possible root cause for the offset error is the non-ideal effect of the silicon beam splitter. Equation~(\ref{eqn_distance}), which is used to generate the estimated distance, is based on the assumption of an ideal beam splitter with zero thickness. However, in a practical setup, the thickness of the silicon beam splitter alters the signal propagation from the transmitter due to refraction. It introduces an additional delay ($\tau_{\mathrm{S}}$) when the signal propagates through, as shown in Fig.~\ref{result_split}. It is observable that the measured total delay in the system ($\tau_\mathrm{m}$) also includes the propagation delay inside the splitter ($\tau_{\mathrm{S}}$) and the corresponding estimated ToF, $\hat{\tau}_{\mathrm{ToF}}$, is

\begin{equation}
    \hat{\tau}_{\mathrm{ToF}} = \tau_{\mathrm{m}} - \tau_{\mathrm{sys}} = \tau_{\mathrm{ToF}} +\tau_{\mathrm{S}} +\tau_{\mathrm{SRA}} - \tau_{\mathrm{SRM}},
\end{equation}
where $\tau_{\mathrm{SRA}}$ and $\tau_{\mathrm{SRM}}$ are the propagation delay from the reflection point of the splitter to the receiver in system delay and target delay measurements, respectively. Thus, it is apparent that the following error factors are introduced due to the thickness of the splitter which causes the estimation offset:
\begin{enumerate}
    \item An additional propagation delay within the beam splitter, $\tau_\mathrm{S}$, which is non-negligible given the refractive index of silicon ($3.42$~\cite{dai2004terahertz}) and the splitter thickness ($3.5$~mm). From the distance detection point of view, $\tau_\mathrm{S}$ should be part of the system delay. However, it cannot be captured with the current system delay measurement approach. For the $3.5$~mm-thick splitter, the offset caused by $\tau_S$ is around $6.11$~mm.
    \item A bias on the measured system internal propagation distance, $\tau_{\mathrm{SRA}} - \tau_{\mathrm{SRM}}$, that is introduced by the mismatch of reflection point locations between the splitter setups for system delay measurement and target distance measurement. This bias does not exist for an ideal splitter with zero thickness as the reflection points are the same (the centre of the splitter). However, for the $3.5$mm-thick splitter, the bias is approximately $-1.952$~mm.
    \item A bias on the ground truth reference distance. The shift of the reflection point also alters the ground truth distance determination. As indicated in Fig.~\ref{result_split}, the ground truth distance should be measured from the reflection point to the target, instead of the centre of the splitter. With the thickness of $3.5$~mm, this bias is approximately $1.75$~mm.
\end{enumerate}

Since the thickness of the beam splitter is consistent, it is possible to directly calibrate the distance estimation result with the offsets mentioned above. Figure~\ref{result_total}(b) demonstrated the effect of calibration with all the offsets compensated. It is observable that offset error has been successfully eliminated, while the calibrated estimates are fairly accurate versus the ground truth distances. Alternatively in practical applications, this offset error can also be calibrated with a simple linear regression method from two or multiple known ground truth points.

\begin{figure}[htbp]
\begin{subfigure}{\textwidth}
\centering\includegraphics[width=0.7\linewidth]{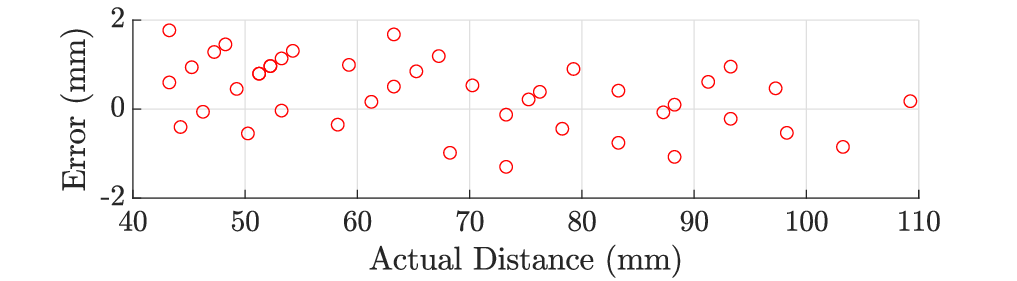}
\caption{ } \label{result_err}
\end{subfigure}
\begin{subfigure}{\textwidth}
\centering\includegraphics[width=0.7\linewidth]{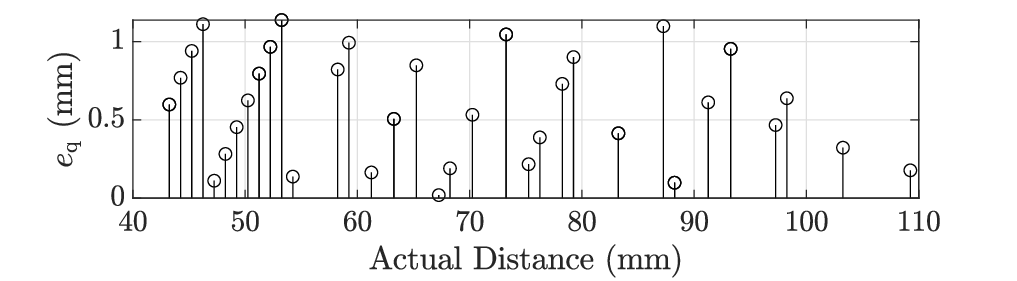}
\caption{ } \label{result_errq}
\end{subfigure}
\begin{subfigure}{\textwidth}
\centering\includegraphics[width=0.7\linewidth]{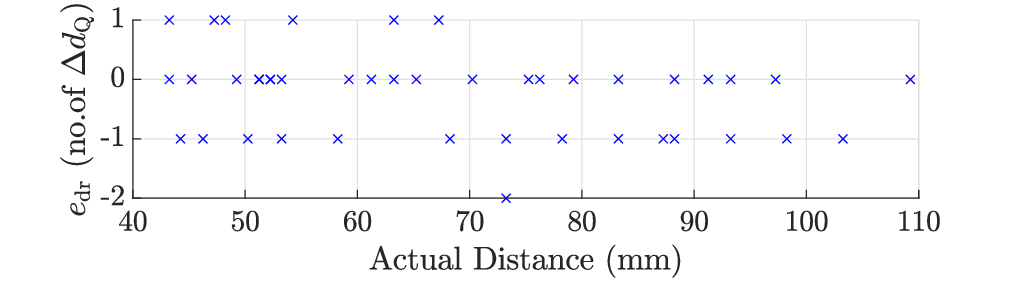}
\caption{ } \label{result_erroffset}
\end{subfigure}
\caption{Insights of the estimation errors remnant after calibration (a) estimation errors for each measurement in mm; (b) the quantization error $e_\mathrm{q}$ in mm and (c) the random drift $e_{\mathrm{dr}}$ in terms of multiples of $\Delta d_{\mathrm{Q}}$.}\label{result_dist_error}
\end{figure}

In addition to the offset error elaborated above, there are other error factors involved, as illustrated in Fig.~\ref{result_dist_error}. Figure~\ref{result_dist_error}(a) indicates the remaining estimation errors after calibration ($e = \hat{d}_{\mathrm{calibrated}} - d$), which can be further decomposed into two components:
\begin{enumerate}
    \item A quantization error $e_\mathrm{q}$ caused by the sampling resolution $\Delta d_\mathrm{Q}$ due to the sampling frequency constraint of the oscilloscope, as discussed in Section~\ref{subsec_dist_setup}. While Fig.~\ref{result_dist_error}(b) shows the respective $e_\mathrm{q}$ for every measurement, it is observable that a quantization error of about $0.17$~mm is accumulated for every $1$~mm increase of target distance, which is aligned with the Eq.~(\ref{eqn_qerr}) as the $\Delta d_\mathrm{Q} \approx 1.17$~mm here.
    
    \item A random drift $e_{\mathrm{dr}}$ as shown in Fig.~\ref{result_dist_error}(c) that varies from  $-\Delta d_\mathrm{Q}$ to $\Delta d_\mathrm{Q}$, with occasional $2\Delta d_\mathrm{Q}$. This is potentially due to the ambiguity where the cross-correlation coefficients of two adjacent sample periods are too close to decide the exact peak location. Various reasons, such as channel and hardware noise can cause this and introduce a typical drift of one $\Delta d_{\mathrm{Q}}$. Since the ToF computation involved both total delay and system delay which are acquired via the same cross-correlation approach, the worst-case drift of $2\Delta d_{\mathrm{Q}}$ can occur. In addition, this error factor is the main cause for the observation that multiple different estimations are reported for different trials with the same ground-truth distance since the deviations between the estimated distances are exactly $\Delta d_{\mathrm{Q}}$.

\end{enumerate}

Both the quantization error and the random drift error mentioned above are inevitable. Thus, they become the limiting factors affecting the distance detection resolution. Next, we will elaborate in detail.

\subsection{Distance Detection Resolution}
\label{subsec_dist_resolution}
\begin{figure}[htbp]
\centering\includegraphics[width=10cm]{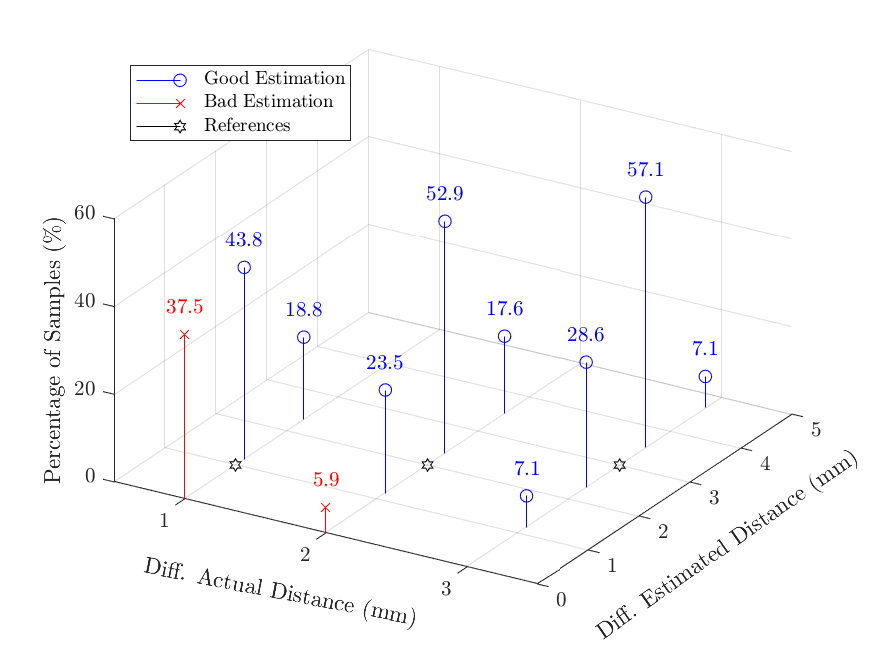}
\caption{Distributions of the difference of estimated distances for measurement pairs with $1$~mm, $2$~mm and $3$~mm actual distance gaps.} \label{result_res}
\end{figure}

We experimentally verified the impact of each error factor on the detection resolution. As discussed previously, it is reasonable to consider the detection resolution on distance in the form of the minimum separation distance between two distinct locations which are distinguishable by the system with adequate accuracy. In other words, the detection is beyond the system resolution if the actual distance difference between two measurements, $\Delta d = d_\mathrm{1} -d_\mathrm{2}$, is too small for the system to adequately distinguish, that is, the system recognizes their estimated distances $\hat{d_\mathrm{1}} = \hat{d_\mathrm{2}}$. To evaluate the resolution of the system, three data groups were formed with pairs of measurements collected of $\Delta d=1$ mm, $2$ mm and $3$ mm respectively. For each measurement pair, the difference in estimated distances, $\Delta \hat{d} = \hat{d_\mathrm{1}} - \hat{d_\mathrm{2}}$, was evaluated as shown in Fig.~\ref{result_res}.

For each group of measurement pairs, the pin locations on the horizontal axis indicate the corresponding $\Delta d$ and $\Delta \hat{d}$, while the height of the pin on the vertical axis shows the percentage of the samples among the respective group. Intuitively, since the two target locations are distinct ($\Delta d>0$) for each measurement pair involved, it is expected that $\Delta \hat{d} >0$ for any successful distinction between the two estimated distance ("Good Estimation"), as shown by blue pins with circles. On the other hand, an unsuccessful distinction can be marked with $\Delta \hat{d} =0$ which contradicts with the ground truth $\Delta d>0$ ("Bad Estimation"), as shown by red pins with crosses. As indicated, for $\Delta d=1$~mm, $37.5$\% of the measurement pairs end up as bad estimation, which is equivalent to a $62.5$\% accuracy in distinguishing two distances with $1$~mm difference; When $\Delta d$ increases to $2$~mm, the accuracy is boosted to $94.1$\%. The accuracy is $100$\% when $\Delta d$ is $3$~mm and above. The observations are aligned with the discussions earlier that the highest achievable resolution is $\Delta d_\mathrm{Q} =1.17$~mm which is the boundary due to the hardware capability. However, the random drift error of another $1.17$~mm due to the ambiguity can occur, which leads to certain estimation errors appearing in $2$~mm resolution evaluation. Thus, empirically it is reasonable to state that the system has a $100$\% accuracy performance in detecting a distance with $3$~mm resolution, with an adequate safe margin. Note that the accuracy performance displayed in Fig.~\ref{result_res} is limited by the number of samples available (about 10-20 samples per group). The exact resolution might be deferred if more data is available or evaluated with a different accuracy threshold to define the resolution.

\section{Conclusion}
We demonstrated a D-band photonics-based communication system designed to investigate wireless communication and integrated localization and sensing. We showed a 5 Gbps communication link for a 1.5 m distance. Atmospheric humidity and the cut-off frequency limit of the system have been experimentally investigated to have minimal effect. The data transfer rate and link distance can be simply improved by using AWG and oscilloscope with higher bandwidth and larger diameter lenses, respectively. The link distance can be also improved by using a terahertz low-noise amplifier with a higher gain. We then utilized the squared-wave signal to measure the target location. We demonstrated that millimetre-order range resolution ($< 3$~mm) can be achieved simultaneously with the same system and discussed the system calibration due to the thickness of the silicon beam splitter and the potential sources of the errors (quantization and random drift errors). This revealed that the error due to the thickness of the beam splitter can be eliminated. While the quantization error and the random drift error are inevitable and are the limiting factors of the resolution achieved.

To conclude, we demonstrate target localization and communication simultaneously, which achieves millimetre-order range resolution. Future works may include an investigation on the relationship between localization performance and communication data rate. This work paves the path towards the implementation of photonics-based D-band integrated localization and communication. Furthermore, this system is easily expendable to any terahertz communication band.

\section{Acknowledgments}
SA and DM acknowledge the partial support by the Australian Government through the Office of National Intelligence’s funded - National Intelligence and Security Discovery Research Grant program. QW acknowledges the support from UNSW Digital Grid Futures Institute's 2024 Seed Funding.

\renewcommand{\bibname}{Reference}
\footnotesize\bibliography{main}
\bibliographystyle{unsrt}

\end{document}